\documentclass[12pt,dvips,oneside,aps,showpacs]{revtex4}
\usepackage{graphicx,xspace,color}
\usepackage[scanall]{psfrag}
\begin{document}
\title{Density effects in a bulk binary Lennard-Jones system} 

\author{Javier Hern\'andez-Rojas$^a$ and David J. Wales$^b$}
\affiliation{$^a$Departamento de F\'\i sica Fundamental II, 
Universidad de La Laguna, 38205 Tenerife, Spain. \\
$^b$University Chemical Laboratories, Lensfield Road, 
Cambridge CB2 1EW, United Kingdom}

\date{\today}

\begin{abstract}
Properties of local minima as a function of density are
studied in a binary Lennard-Jones system for
kinetic equipartition temperatures $T=1.0$ (normal liquid), 
$0.5$ (supercooled liquid), and $0.4$ (glass), 
in reduced units. The number of
different local minima sampled, energy, pressure, normal mode
angular frequencies, mean distance between all pairs 
of local minima and partial radial distribution
functions are presented. In agreement with previous studies by 
Sastry [Phys.~Rev.~Lett.~{\bf 85}, 590 (2000)] a limiting density 
is found at $\rho_l=1.06$ with negative pressure, 
below which the local structure of the glass 
and the supercooled phases are essentially the same, 
as evidenced by the partial radial 
distribution functions. 
The mean energy, pressure, and normal mode frequencies 
have values that are practically independent of the 
temperature below $\rho_l$. 
\end{abstract}
\vfill

\maketitle

\section{Introduction}
In recent years a considerable research effort has been expended to 
understand the complex phenomenology of supercooled liquids and glasses. 
Mode-coupling theory has proved quite successful for supercooled liquids
at higher temperatures, before activated dynamics must be accounted
for explicitly.\cite{Got91,Got92,Kob97,Got99}
It was Goldstein who first related the behaviour of 
glass formers to the underlying potential energy surface (PES). 
\cite{Gol69} In this approach, the dynamics are separated into vibrational
motion about a minimum on the PES and transitions between local minima
or `inherent structures'.\cite{Still82,Still84} 
The PES is partitioned into basins of attraction surrounding the 
local minima, where a basin of attraction is defined as a set of points
that lead to the same minimum along steepest-descent pathways.
Recently, it has been possible to establish connections 
between the structure, dynamics and thermodynamics of finite systems and the
PES in some detail.\cite{WalesDMMW00}

Following Angell,\cite{Angell95,Angell97,Angell99} 
glass-forming liquids can be classified as strong
or fragile. Fragile systems exhibit non-Arrhenius temperature dependence
of transport properties such as the diffusion constant, usually accompanied
by a significant heat capacity peak at the glass transition. In contrast, 
strong systems exhibit Arrhenius dynamics and small or negligible changes 
in the thermodynamic properties at the glass transition. 

Computer simulation based on standard molecular 
dynamics (MD) or Monte Carlo (MC) techniques, 
has played a significant role in the 
testing and development of models for supercooled liquids and
glasses.\cite{Kob99} Interest in more direct connections to the PES has
recently increased, with Sastry, Debenedetti and Stillinger\cite{Still98} 
characterising `landscape-influenced' and `landscape-dominated'
regimes for a binary Lennard-Jones system, in agreement with
the instantaneous normal modes picture of Donati, Sciortino and
Tartaglia.\cite{Donati00} Recently, a kinetic Monte Carlo (KMC) 
approach\cite{Voter86,Wein91,Jonsson01} was used to
provide an alternative view of the dynamics.\cite{Javier02} 
   
In the present contribution we focus on some properties of the energy 
landscape in a binary Lennard-Jones (BLJ) mixture, namely, the effect
of the density on the local minima sampled as a function of temperature.
This topic has been previously considered by 
Malandro et al.\cite{Lacks97,Lacks98} and Sastry.\cite{Sastry00,Sastry01} 
In the former studies the volume dependence
of the local minima appearances and disappearances were examined. Similar 
results were found by Heuer\cite{Heuer97}.
Sastry also
analysed the relationship between fragility and the PES and how the former 
depends on the bulk density. One feature
of the BLJ system is that the vibrational contribution to the entropy, 
within the 
harmonic approximation, decreases with the potential energy. 
Furthermore, Sastry 
found a limiting density,
$\rho_l=1.08$ in reduced units, which defines a limit 
of stability separating spatially heterogeneous 
structures (below $\rho_l$) from
more homogeneous structures (above $\rho_l$). $\rho_l$ was 
also interpreted as a density limit to glass formation.           
Here, we concentrate on this limit and consider 
the glass and supercooled liquid structures in more detail 
as a function of density. 


\section{Methods}

The model we used is a binary mixture of $N=256$ atoms 
in a cubic box with periodic boundary conditions, 
containing 205 ($\sim 80\%$) A atoms and 51 ($\sim 20\%$)
B atoms interacting according to a Lennard-Jones pair potential of
the form
\begin{equation}
V_{\alpha \beta}=4\epsilon_{\alpha \beta}
\left[ \left( \frac{\sigma_{\alpha \beta}}{r_{\alpha \beta}} \right)^{12}-
\left( \frac{\sigma_{\alpha \beta}}{r_{\alpha \beta}} \right)^{6} \right],
\end{equation}
$r_{\alpha \beta}$ being the distance between particles $\alpha$ and 
$\beta$. The values of 
the Lennard-Jones parameters are $\epsilon_{AA}=1.0$, $\epsilon_{BB}=0.5$,
$\epsilon_{AB}=1.5$, $\sigma_{AA}=1.0$, $\sigma_{BB}=0.88$ and 
$\sigma_{AB}=0.8$.\cite{Kob94}
The units of distance, energy, temperature, pressure, and time were taken as
$\sigma_{AA}$, $\epsilon_{AA}$, $\epsilon_{AA}/k_B$ ($k_B$ is the 
Boltzmann constant), $\epsilon_{AA} \sigma_{AA}^{-3}$ 
and $\sigma_{AA}(m/\epsilon_{AA})^{1/2}$, with 
$m$ the mass of both A and B atoms. The initial density was 
$1.2 \sigma_{AA}^{-3}$ with a fixed cutoff of $2.5 
\sigma_{\alpha \beta}$
along with the minimum image convention. We truncated and shifted 
the potential with a quadratic function, so that the energy and its 
first derivative are continuous at the cutoff value.\cite{Still98,Ford73} 

This BLJ model has been extensively studied 
in the glasses community, as it does not crystallise 
on the molecular dynamics 
time scale.\cite{Kob99,Still98,Donati00,Javier02,Sastry00,Sastry01,Kob94,Tom01,Mousseau,Angelani00,Broderix00,Schroder00,2Sastry00,Jonsson88,Kob95,2Kob95,Donati99,Kob00,2Angell99,Heuer00,Bagchi02,Jon02}
However, we have recently found that there are crystalline minima for 
this system, based on face-centred-cubic A atoms with trigonal or square
prismatic B-A coordination.\cite{2Tom01}
Previous work has
shown that this system exhibits a significant degree
of non-Arrhenius behaviour at low temperatures, i.e. it is 
fragile in Angell's terminology.\cite{Kob94,Kob95,2Kob95,Donati99}

Microcanonical molecular dynamics simulations were carried out, 
where the classical equations of motion were integrated using a Verlet 
algorithm. We employed 10$^5$ equilibration steps, followed by 
10$^6$ data collection steps with a time step of 0.003 in reduced 
units.\cite{Jon02}  

In order to study the local minima 
the instantaneous system configuration was quenched, every 1000th
configuration, to locate a minimum. Quenching was performed
using a modified version of Nocedal's limited 
memory Broyden-Fletcher-Goldfarb-Shanno (L-BFGS) 
algorithm.\cite{WalesDMMW00,Nocedal89,Wales99}
Each quench was finished using a few eigenvector-following steps\cite{Miller81,Wales94} 
to converge the
root-mean-square gradient below $10^{-7}$ reduced units, employing full
diagonalisation of the analytical Hessian matrix to ensure that the
stationary points have the correct Hessian index (the number of
negative eigenvalues). 
In this manner we generated a sample of minima
at density $\rho=1.2$.  

For each local minimum, the density of the system
was changed by 0.02 in reduced units. To do this we changed 
the box length to give the density required, 
without rescaling the coordinates of the atoms in the box.
Then, the potential energy
was reminimized using the methods described
above. This process was repeated in order to follow the 
properties of the minima over a wide
range of density.

We have analysed the behaviour of the local minima at
densities in the range $0.6\leq\rho\leq1.3$, and 
for three different initial kinetic equipartition temperatures 
$1.0$, $0.5$ and $0.4$. It is known for this 
system that the glass transition temperature, $T_g$, 
occurs between 0.5 
and 0.4.\cite{Still98,Javier02,Jonsson88,Jon02}

\section{Results}
In this section we present the results obtained for 
the behaviour of 
the local minima at different temperatures and densities.
At the initial density $\rho=1.2$, the number of different local minima
that we found from the 1000 instantaneous configurations, 
$N_{min}$, was $1000$, $991$, and $268$ at  
$T=1.0$, $0.5$, and $0.4$, respectively. 
These values reflect the increasing residence times in the 
local minima as the temperature is decreased, with a dramatic
fall in $N_{min}$ below $T_g$.
We found that $N_{min}$ 
is basically independent of the density in 
the range of densities studied.

\begin{figure}[hp]
\psfrag{-5.8}[cr][cr]{\Large $-5.8$}
\psfrag{-6.0}[cr][cr]{\Large $-6.0$}
\psfrag{-6.2}[cr][cr]{\Large $-6.2$}
\psfrag{-6.4}[cr][cr]{\Large $-6.4$}
\psfrag{-6.6}[cr][cr]{\Large $-6.6$}
\psfrag{-6.8}[cr][cr]{\Large $-6.8$}
\psfrag{-7.0}[cr][cr]{\Large $-7.0$}
\psfrag{-7.2}[cr][cr]{\Large $-7.2$}
\psfrag{0.6}[tc][tc]{\Large $0.6$}
\psfrag{0.7}[tc][tc]{\Large $0.7$}
\psfrag{0.8}[tc][tc]{\Large $0.8$}
\psfrag{0.9}[tc][tc]{\Large $0.9$}
\psfrag{1.0}[tc][tc]{\Large $1.0$}
\psfrag{1.1}[tc][tc]{\Large $1.1$}
\psfrag{1.2}[tc][tc]{\Large $1.2$}
\psfrag{1.3}[tc][tc]{\Large $1.3$}
\psfrag{T=1}[Bl][Bl]{$T=1.0$}
\psfrag{T=0.5}[Bl][Bl]{$T=0.5$}
\psfrag{T=0.4}[Bl][Bl]{$T=0.4$}
\psfrag{Energy}[Bc][tc]{\Large mean potential energy}
\psfrag{Density}[tc][tc]{\Large $\rho$}
\psfrag{9}[cr][cr]{\Large $9$}
\psfrag{6}[cr][cr]{\Large $6$}
\psfrag{3}[cr][cr]{\Large $3$}
\psfrag{0}[cr][cr]{\Large $0$}
\psfrag{-3}[cr][cr]{\Large $-3$}
\psfrag{-6}[cr][cr]{\Large $-6$}
\psfrag{0.6}[tc][tc]{\Large $0.6$}
\psfrag{0.7}[tc][tc]{\Large $0.7$}
\psfrag{0.8}[tc][tc]{\Large $0.8$}
\psfrag{0.9}[tc][tc]{\Large $0.9$}
\psfrag{1.0}[tc][tc]{\Large $1.0$}
\psfrag{1.1}[tc][tc]{\Large $1.1$}
\psfrag{1.2}[tc][tc]{\Large $1.2$}
\psfrag{1.3}[tc][tc]{\Large $1.3$}
\psfrag{T=1}[Bl][Bl]{$T=1.0$}
\psfrag{T=0.5}[Bl][Bl]{$T=0.5$}
\psfrag{T=0.4}[Bl][Bl]{$T=0.4$}
\psfrag{Pressure}[Bc][tc]{\Large $P$}
\psfrag{Density}[tc][tc]{\Large $\rho$}
\centerline{
\includegraphics[width=14cm]{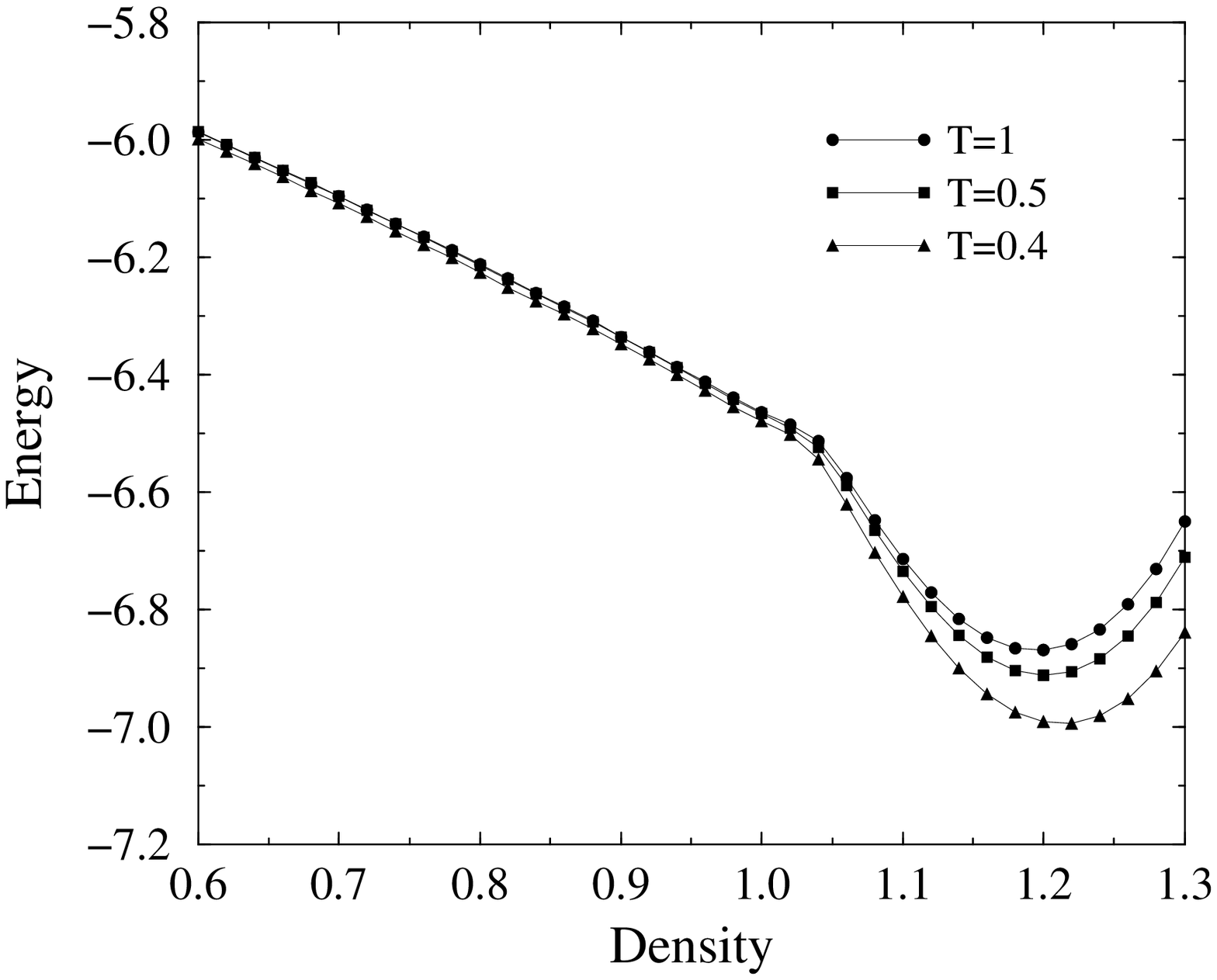}}
\centerline{
\includegraphics[width=14cm]{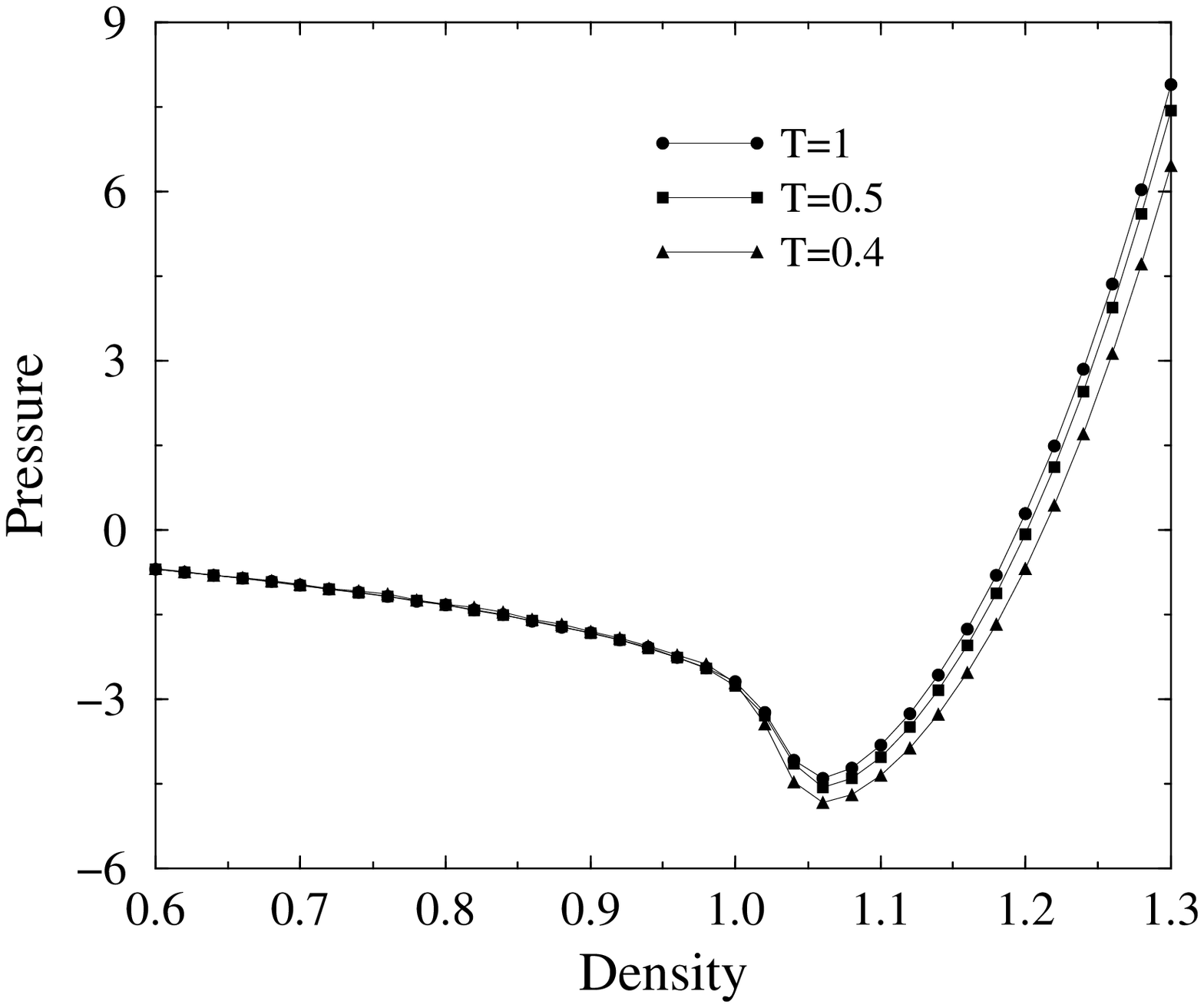}}
\caption{Mean potential energy (top) and pressure (bottom) as a function of the density 
for the three reoptimised samples of local minima.}
\label{fig:energy}
\end{figure}

We have also evaluated the mean potential energy for the 
local minima as a function of density for
three different initial samples at $T=1.0$, 0.5 and 0.4.
In Fig.~\ref{fig:energy} we can see how the lowest value of 
the energy occurs around $\rho=1.2$
for each temperature. 
However, at $T=0.4$ the minimum of the curve is slightly 
displaced to higher density, namely, to $\rho=1.22$. 

A clear change of slope appears in Fig.~\ref{fig:energy} 
(the system is mechanically unstable) at 
$\rho_l=1.06$ for each temperature, and for
$\rho<\rho_l$
the mean potential energy
is practically independent of $T$ and increases as the density 
decreases. 


Another interesting property is the variation of 
the pressure with density (Fig.~\ref{fig:energy}). 
In this case, the pressure, $P$, was calculated
using the virial equation\cite{Allen87}

\begin{equation}
PV=(N-1)T+\frac{1}{3}<\sum_i \sum_{j>i} {\bf r}_{ij} \cdot {\bf F}_{ij}>,
\end{equation}
where $V=N/\rho$ is the volume of the supercell, 
$N$ is the number of atoms,
${\bf r}_{ij}={\bf r}_i-{\bf r}_j$ and ${\bf F}_{ij}$ 
is the force on the {\it i}th particle due to {\it j}th particle. 
A factor $N-1$ appears because of momentum conservation.
We used $T=0$ to calculate the pressure for the local minima. 

The lowest value of $P$ is reached for each temperature 
around $\rho_l=1.06$. This value is very close to Sastry's result of 
$\rho_l=1.08$\cite{Sastry00} (his supercell contained 204 A atoms and
52 B atoms). 
For $\rho<1$, the mean pressure is practically independent of 
the temperature. 
Fig.~\ref{fig:energy} also shows that the pressure 
is nearly zero at densities corresponding to the lowest 
values of the mean energy.

\begin{figure}[hp]
\psfrag{20}[cr][cr]{\Large 20}
\psfrag{18}[cr][cr]{\Large 18}
\psfrag{16}[cr][cr]{\Large 16}
\psfrag{14}[cr][cr]{\Large 14}
\psfrag{12}[cr][cr]{\Large 12}
\psfrag{10}[cr][cr]{\Large 10}
\psfrag{0.6}[tc][tc]{\Large 0.6}
\psfrag{0.7}[tc][tc]{\Large 0.7}
\psfrag{0.8}[tc][tc]{\Large 0.8}
\psfrag{0.9}[tc][tc]{\Large 0.9}
\psfrag{1.0}[tc][tc]{\Large 1.0}
\psfrag{1.1}[tc][tc]{\Large 1.1}
\psfrag{1.2}[tc][tc]{\Large 1.2}
\psfrag{1.3}[tc][tc]{\Large 1.3}
\psfrag{T=1}[Bl][Bl]{$T=1$}
\psfrag{T=0.5}[Bl][Bl]{$T=0.5$}
\psfrag{T=0.4}[Bl][Bl]{$T=0.4$}
\psfrag{w}[Bc][tc]{\Large $<\omega>$}
\psfrag{Density}[tc][tc]{\Large $\rho$}
\psfrag{5}[cr][cr]{\Large 5}
\psfrag{4}[cr][cr]{\Large 4}
\psfrag{3}[cr][cr]{\Large 3}
\psfrag{2}[cr][cr]{\Large 2}
\psfrag{1}[cr][cr]{\Large 1}
\psfrag{0}[cr][cr]{\Large 0}
\psfrag{0.6}[tc][tc]{\Large 0.6}
\psfrag{0.7}[tc][tc]{\Large 0.7}
\psfrag{0.8}[tc][tc]{\Large 0.8}
\psfrag{0.9}[tc][tc]{\Large 0.9}
\psfrag{1.0}[tc][tc]{\Large 1.0}
\psfrag{1.1}[tc][tc]{\Large 1.1}
\psfrag{1.2}[tc][tc]{\Large 1.2}
\psfrag{1.3}[tc][tc]{\Large 1.3}
\psfrag{T=1}[Bl][Bl]{$T=1.0$}
\psfrag{T=0.5}[Bl][Bl]{$T=0.5$}
\psfrag{T=0.4}[Bl][Bl]{$T=0.4$}
\psfrag{dij}[Bc][tc]{\Large $\left<d_{\rm ij}\right>$}
\psfrag{Density}[tc][tc]{\Large $\rho$}
\centerline{
\includegraphics[width=14cm]{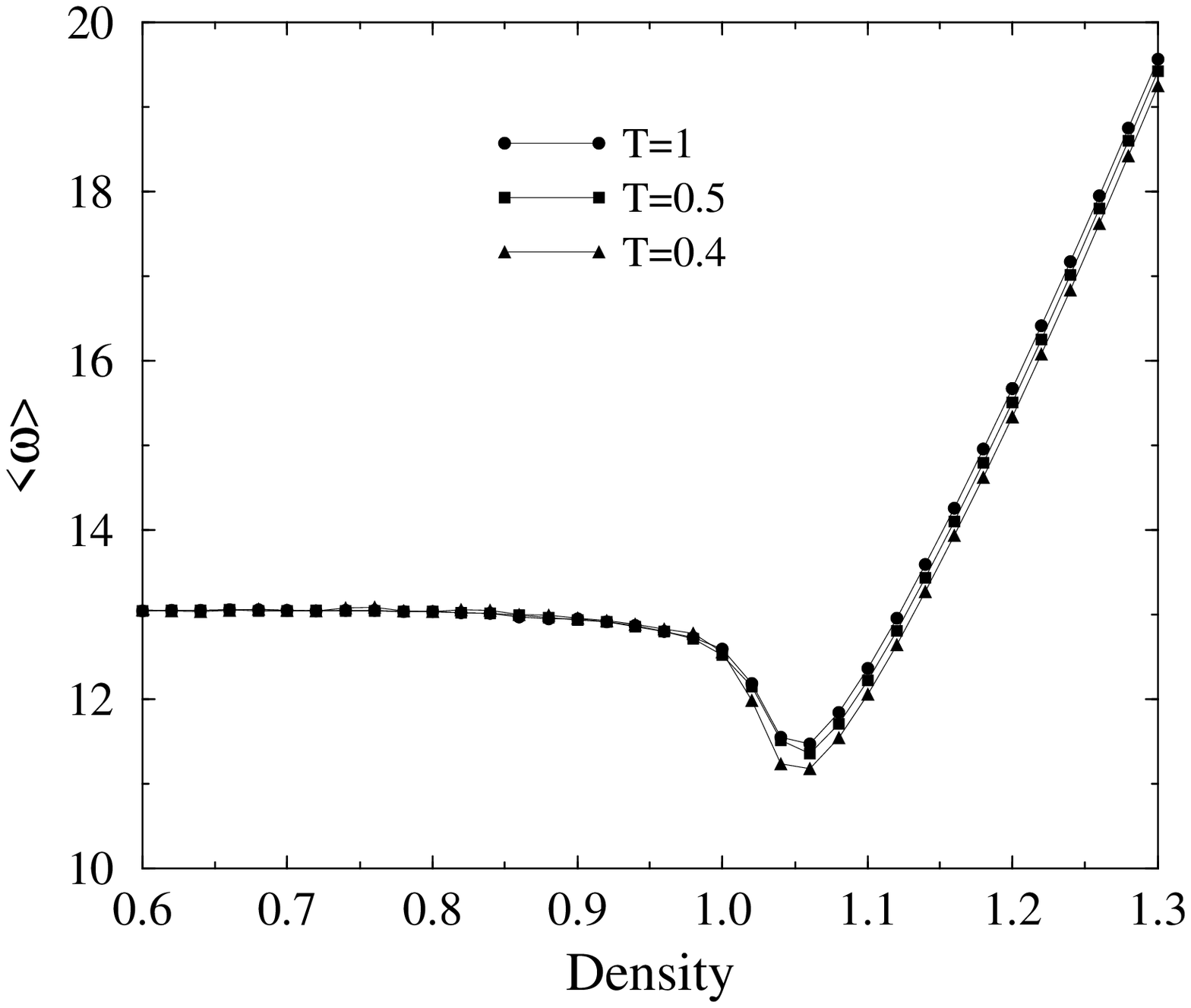}}
\centerline{
\includegraphics[width=14cm]{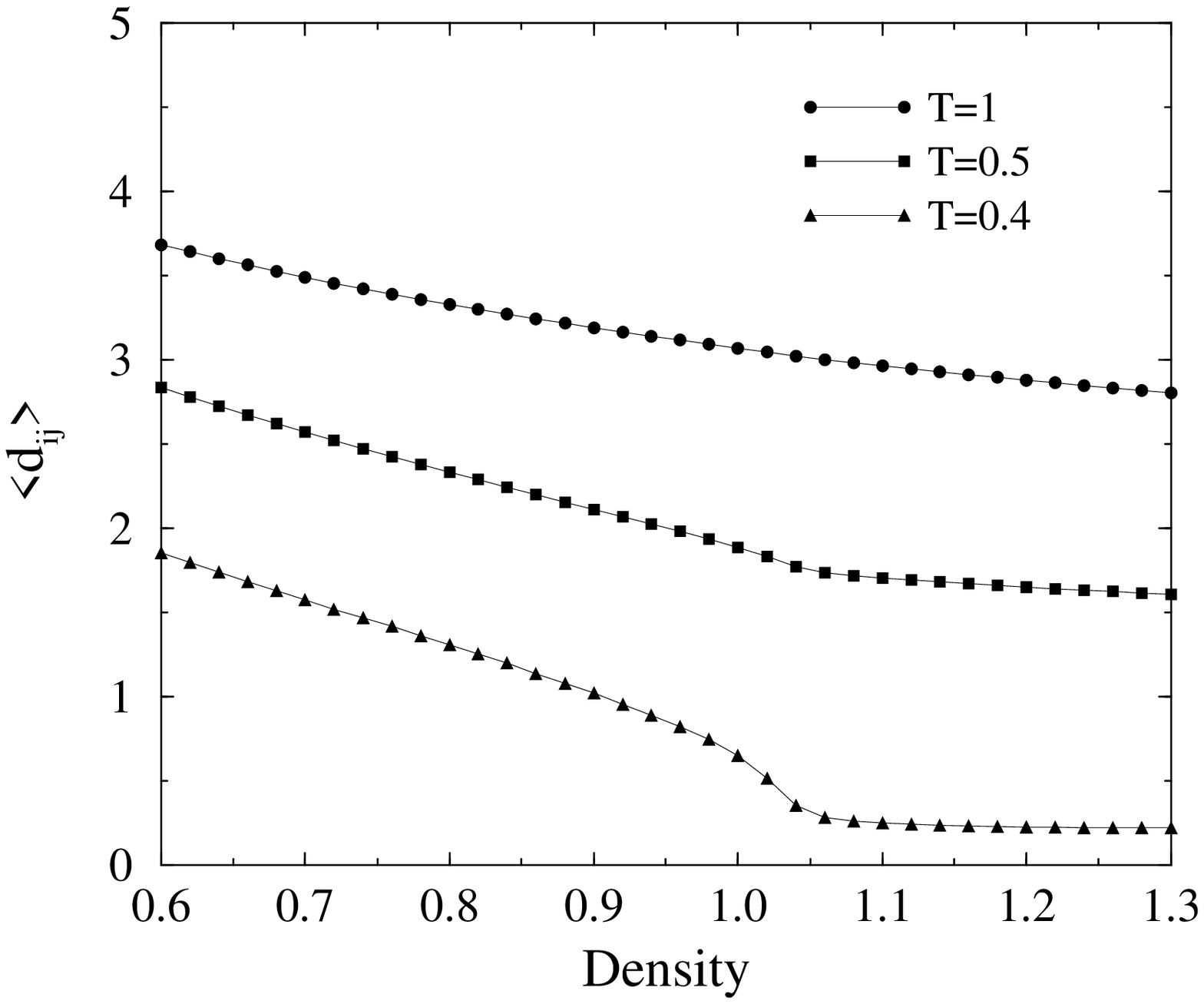}}
\caption{Mean angular frequency (top) and distance between all pairs of local minima (bottom) 
as a function of the density for the three samples of reoptimised local minima.}
\label{fig:freqdist}
\end{figure}

The geometric mean normal mode frequency at 
each local minimum is 

\begin{equation}
\overline{\nu}=\prod_{i=1}^{3N-3} (\nu_i)^{1/(3N-3)},
\end{equation}
where the $\nu_i$ are obtained by diagonalising the Hessian matrix. 
A mean angular frequency, averaged over all the local minima, can be 
defined as 

\begin{equation}
<\omega>=2\pi<\overline{\nu}>,
\end{equation}
and this quantity is plotted in Fig.~\ref{fig:freqdist}. 
The frequencies are important for dynamics because they
appear in the usual transition state theory rate constant for barrier
crossing. If we also use the 
information obtained in Fig.~\ref{fig:energy} 
we can see that for $\rho>1.2$, 
which corresponds to positive pressures (compressed regime),
$<\omega>$ increases with
energy, in agreement with Sastry.\cite{Sastry01} 
However, between $\rho\sim 1.2$ and $\rho_l=1.06$, which corresponds 
to negative pressures (stretched regime), 
we see the opposite behaviour, i.e.~$<\omega>$ decreases as the 
density decreases. Then, it increases again to reach a constant value
of 13 below $\rho\sim 1.0$ independent of $T$.   
 


We have also calculated the mean value of the distance
between all pairs of local minima sampled:
 
\begin{equation}
d_{ij}=\sqrt{\frac{1}{N} \sum_{k=1}^N \left| {\bf r}_k^i
-{\bf r}_k^j \right| ^2},
\end{equation}
where $i$ and $j$ indicates the $i$th and $j$th local minima. It is 
clear from Fig.~\ref{fig:freqdist} that this quantity 
increases with the initial temperature used
to generate the samples and with decreasing 
density. 
On the other hand, the initial MD trajectories sample 
the PES less extensively at low temperature for 
the same simulation time. In the 
regime $\rho_l=1.06\leq \rho \leq 1.3$ $<d_{ij}>$ does 
not change much, while below 
$\rho_l=1.06$ it increases steadily.

\begin{figure}[hp]
\psfrag{12}[cr][cr]{\Large 12}
\psfrag{8}[cr][cr]{\Large 8}
\psfrag{4}[cr][cr]{\Large 4}
\psfrag{0}[cr][cr]{\Large 0}
\psfrag{0.5}[tc][tc]{\Large 0.5}
\psfrag{1.0}[tc][tc]{\Large 1.0}
\psfrag{1.5}[tc][tc]{\Large 1.5}
\psfrag{2.0}[tc][tc]{\Large 2.0}
\psfrag{2.5}[tc][tc]{\Large 2.5}
\psfrag{r=0.60}[Bl][Bl]{$\rho=0.60$}
\psfrag{r=1.00}[Bl][Bl]{$\rho=1.00$}
\psfrag{r=1.02}[Bl][Bl]{$\rho=1.02$}
\psfrag{r=1.04}[Bl][Bl]{$\rho=1.04$}
\psfrag{r=1.06}[Bl][Bl]{$\rho=1.06$}
\psfrag{r=1.08}[Bl][Bl]{$\rho=1.08$}
\psfrag{r=1.20}[Bl][Bl]{$\rho=1.20$}
\psfrag{r=1.30}[Bl][Bl]{$\rho=1.30$}
\psfrag{r}[tc][tc]{\Large $r$}
\psfrag{gAA}[Bc][tc]{\Large $g_{\rm AA}(r)$}
\psfrag{40}[cr][cr]{\Large 40}
\psfrag{35}[cr][cr]{\Large 35}
\psfrag{30}[cr][cr]{\Large 30}
\psfrag{25}[cr][cr]{\Large 25}
\psfrag{20}[cr][cr]{\Large 20}
\psfrag{15}[cr][cr]{\Large 15}
\psfrag{10}[cr][cr]{\Large 10}
\psfrag{5}[cr][cr]{\Large 5}
\psfrag{0}[cr][cr]{\Large 0}
\psfrag{0.5}[tc][tc]{\Large 0.5}
\psfrag{1.0}[tc][tc]{\Large 1.0}
\psfrag{1.5}[tc][tc]{\Large 1.5}
\psfrag{2.0}[tc][tc]{\Large 2.0}
\psfrag{2.5}[tc][tc]{\Large 2.5}
\psfrag{r=0.60}[Bl][Bl]{$\rho=0.60$}
\psfrag{r=1.00}[Bl][Bl]{$\rho=1.00$}
\psfrag{r=1.02}[Bl][Bl]{$\rho=1.02$}
\psfrag{r=1.04}[Bl][Bl]{$\rho=1.04$}
\psfrag{r=1.06}[Bl][Bl]{$\rho=1.06$}
\psfrag{r=1.08}[Bl][Bl]{$\rho=1.08$}
\psfrag{r=1.20}[Bl][Bl]{$\rho=1.20$}
\psfrag{r=1.30}[Bl][Bl]{$\rho=1.30$}
\psfrag{gAB}[Bc][tc]{\Large $g_{\rm AB}(r)$}
\psfrag{r}[tc][tc]{\Large $r$}
\centerline{
\includegraphics[width=13.5cm]{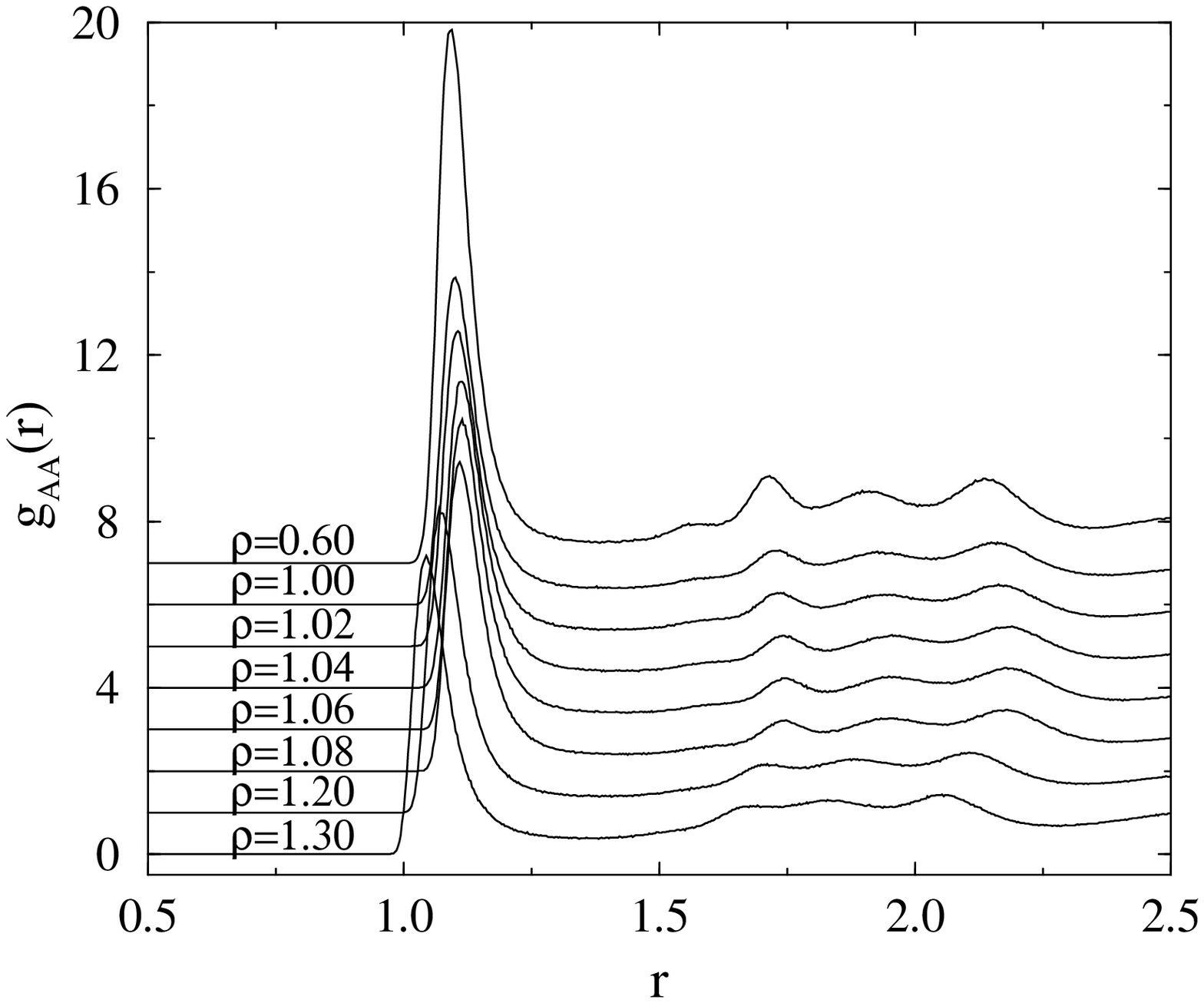}}
\centerline{
\includegraphics[width=13.5cm]{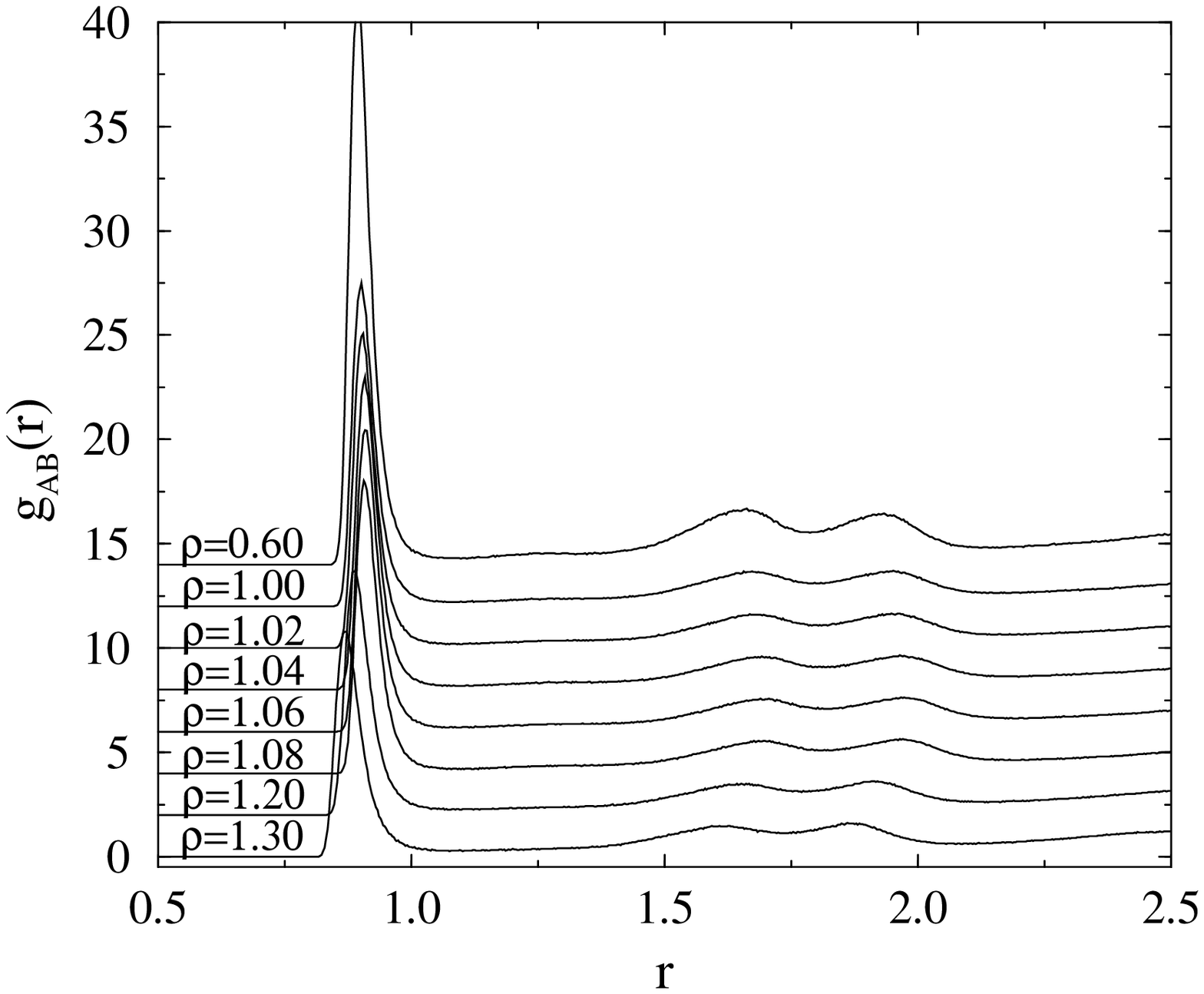}}
\caption{Partial radial distribution function for A particles (top)
and A and B particles (bottom)
averaged over reoptimised local minima initially sampled at $T=1$.
For clarity, the curves have been displaced vertically.}
\label{fig:rdfA}
\end{figure}

The above results were checked 
by studying the partial 
radial distribution functions (PRDF) for all the 
local minima sampled,\cite{Kob95}

\begin{equation}
g_{\alpha \alpha}(r)=\frac{V}{N_{\alpha}(N_{\alpha}-1)}
\left< \sum_{i}^{N_\alpha} \sum_{j\neq i}^{N_\alpha} 
\delta (r-\left| {\bf r}_{ij} \right|) \right>
\end{equation}
and
\begin{equation}
g_{AB}(r)=\frac{V}{N_A N_B}
\left< \sum_{i}^{N_A} \sum_{j}^{N_B} 
\delta (r-\left| {\bf r}_{ij} \right|) \right>,
\end{equation}
where $\delta(r)$ is the delta function. 
In Fig.~\ref{fig:rdfA} we show the PRDF 
for A particles as a function of the density 
for reoptimised local minima generated from the run at $T=1$. 

The first peak is the highest in the PRDF for 
all densities, indicating that the AA correlation takes place at the first coordination
shell. However, as the density is reduced from $\rho=1.3$, 
this peak is slightly displaced to longer distance, i.e. it 
reflects the lower density, and then reaches a 
maximum value at density $\rho_l=1.06$. Below this density, the system shows
the opposite trend due to the formation of 
voids and fractures,\cite{Sastry00,Sastry97,Still00,Corti97}, allowing 
the AA distances to relax back closer to their ideal value.  

\begin{figure}
\begin{psfrags}
\psfrag{8}[cr][cr]{\Large 8}
\psfrag{6}[cr][cr]{\Large 6}
\psfrag{4}[cr][cr]{\Large 4}
\psfrag{2}[cr][cr]{\Large 2}
\psfrag{0}[cr][cr]{\Large 0}
\psfrag{0.5}[tc][tc]{\Large 0.5}
\psfrag{1.0}[tc][tc]{\Large 1.0}
\psfrag{1.5}[tc][tc]{\Large 1.5}
\psfrag{2.0}[tc][tc]{\Large 2.0}
\psfrag{2.5}[tc][tc]{\Large 2.5}
\psfrag{r=0.60}[Bl][Bl]{$\rho=0.60$}
\psfrag{r=1.00}[Bl][Bl]{$\rho=1.00$}
\psfrag{r=1.02}[Bl][Bl]{$\rho=1.02$}
\psfrag{r=1.04}[Bl][Bl]{$\rho=1.04$}
\psfrag{r=1.06}[Bl][Bl]{$\rho=1.06$}
\psfrag{r=1.08}[Bl][Bl]{$\rho=1.08$}
\psfrag{r=1.20}[Bl][Bl]{$\rho=1.20$}
\psfrag{r=1.30}[Bl][Bl]{$\rho=1.30$}
\psfrag{gBB}[Bc][tc]{\Large $g_{\rm BB}(r)$}
\psfrag{r}[tc][tc]{\Large $r$}
\centerline{
\includegraphics[width=14cm]{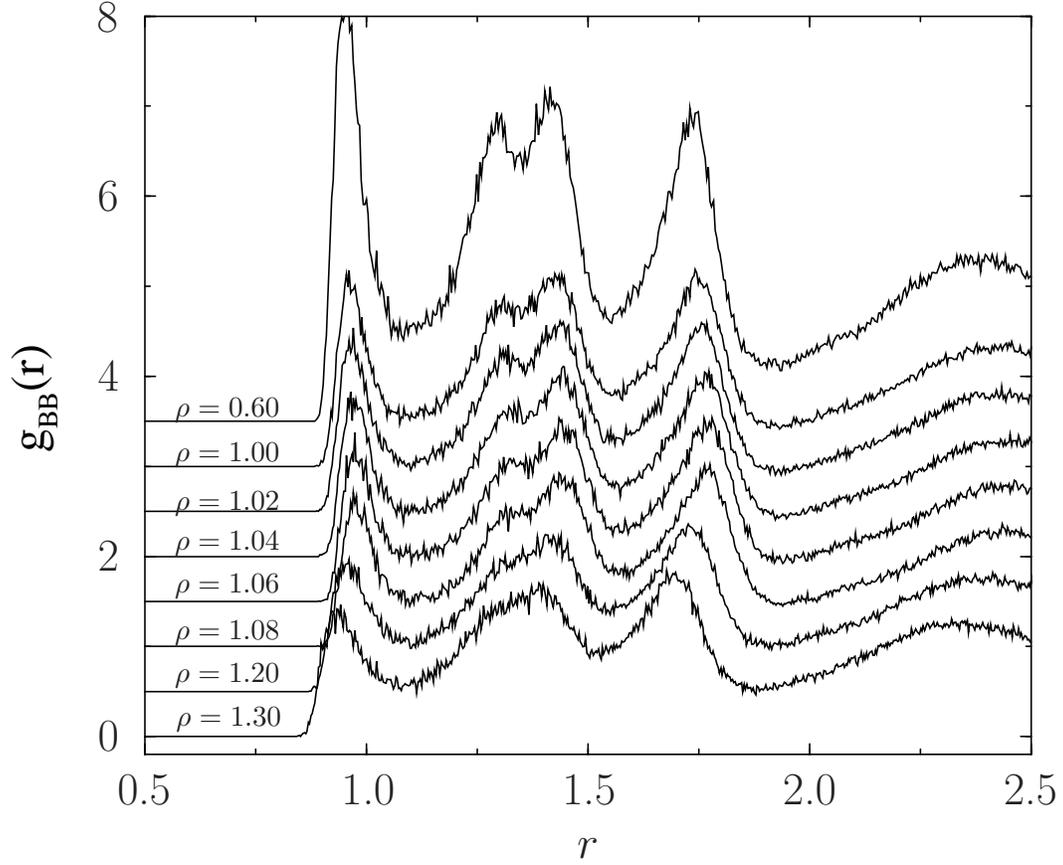}}
\end{psfrags}
\caption{Partial radial distribution function for B particles 
averaged over the reoptimised local minima initially sampled
at $T=1$. As the number 
of B particles is relatively small, $g_{BB}$ is somewhat noisy.
For clarity, the curves have been displaced vertically.}
\label{fig:rdfB}
\end{figure}

Similar behaviour is exhibited in $g_{AB}$ in Fig.~\ref{fig:rdfA}. The strong
attractive interaction between A and B particles is reflected in the 
first peak.
However, $g_{BB}$ in Fig.~\ref{fig:rdfB} exhibits some 
new features. The weaker attractive interaction between B particles is 
reflected in the intensities of the peaks. 
The BB correlation takes place
mainly in the first, second and third coordination shells, but  
for $\rho=0.6$ the first coordination shell is more 
important. Furthermore, a clear splitting appears in the second peak, 
indicating that the local structure for B particles is more ordered 
at low density. 
This is because at low density, with the formation of voids 
in the structure, a dense packing appears. A similar signature
is known at high pressure.\cite{Bagchi02,Wah91}   

\begin{figure}[hp]
\psfrag{14}[cr][cr]{\Large 14}
\psfrag{12}[cr][cr]{\Large 12}
\psfrag{10}[cr][cr]{\Large 10}
\psfrag{8}[cr][cr]{\Large 8}
\psfrag{6}[cr][cr]{\Large 6}
\psfrag{4}[cr][cr]{\Large 4}
\psfrag{2}[cr][cr]{\Large 2}
\psfrag{0}[cr][cr]{\Large 0}
\psfrag{0.5}[tc][tc]{\Large 0.5}
\psfrag{1.0}[tc][tc]{\Large 1.0}
\psfrag{1.5}[tc][tc]{\Large 1.5}
\psfrag{2.0}[tc][tc]{\Large 2.0}
\psfrag{2.5}[tc][tc]{\Large 2.5}
\psfrag{T=0.5}[Bl][Bl]{$T=0.5$}
\psfrag{T=0.4}[Bl][Bl]{$T=0.4$}
\psfrag{r=0.60}[Bl][Bl]{$\rho=0.60$}
\psfrag{r=1.00}[Bl][Bl]{$\rho=1.00$}
\psfrag{r=1.06}[Bl][Bl]{$\rho=1.06$}
\psfrag{r=1.08}[Bl][Bl]{$\rho=1.08$}
\psfrag{r=1.20}[Bl][Bl]{$\rho=1.20$}
\psfrag{r=1.30}[Bl][Bl]{$\rho=1.30$}
\psfrag{gAA}[Bc][tc]{\Large $g_{\rm AA}(r)$}
\psfrag{r}[tc][tc]{\Large $r$}
\psfrag{40}[cr][cr]{\Large 40}
\psfrag{36}[cr][cr]{\Large 36}
\psfrag{32}[cr][cr]{\Large 32}
\psfrag{28}[cr][cr]{\Large 28}
\psfrag{24}[cr][cr]{\Large 24}
\psfrag{20}[cr][cr]{\Large 20}
\psfrag{16}[cr][cr]{\Large 16}
\psfrag{12}[cr][cr]{\Large 12}
\psfrag{8}[cr][cr]{\Large 8}
\psfrag{4}[cr][cr]{\Large 4}
\psfrag{0}[cr][cr]{\Large 0}
\psfrag{0.5}[tc][tc]{\Large 0.5}
\psfrag{1.0}[tc][tc]{\Large 1.0}
\psfrag{1.5}[tc][tc]{\Large 1.5}
\psfrag{2.0}[tc][tc]{\Large 2.0}
\psfrag{2.5}[tc][tc]{\Large 2.5}
\psfrag{T=0.5}[Bl][Bl]{$T=0.5$}
\psfrag{T=0.4}[Bl][Bl]{$T=0.4$}
\psfrag{r=0.60}[Bl][Bl]{$\rho=0.60$}
\psfrag{r=1.00}[Bl][Bl]{$\rho=1.00$}
\psfrag{r=1.06}[Bl][Bl]{$\rho=1.06$}
\psfrag{r=1.08}[Bl][Bl]{$\rho=1.08$}
\psfrag{r=1.20}[Bl][Bl]{$\rho=1.20$}
\psfrag{r=1.30}[Bl][Bl]{$\rho=1.30$}
\psfrag{gAB}[Bc][tc]{\Large $g_{\rm AB}(r)$}
\psfrag{r}[tc][tc]{\Large $r$}
\centerline{
\includegraphics[width=14cm]{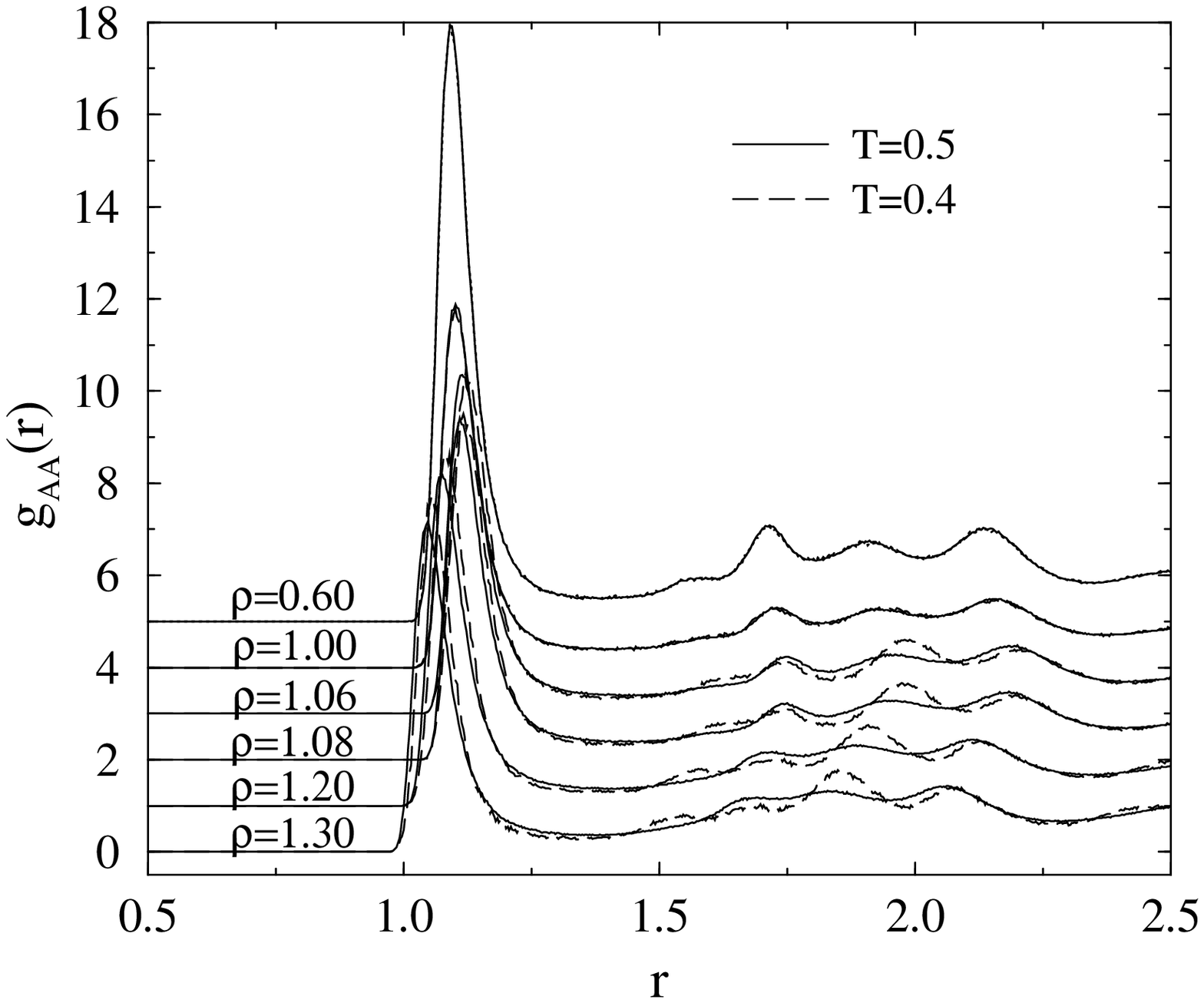}}
\centerline{
\includegraphics[width=14cm]{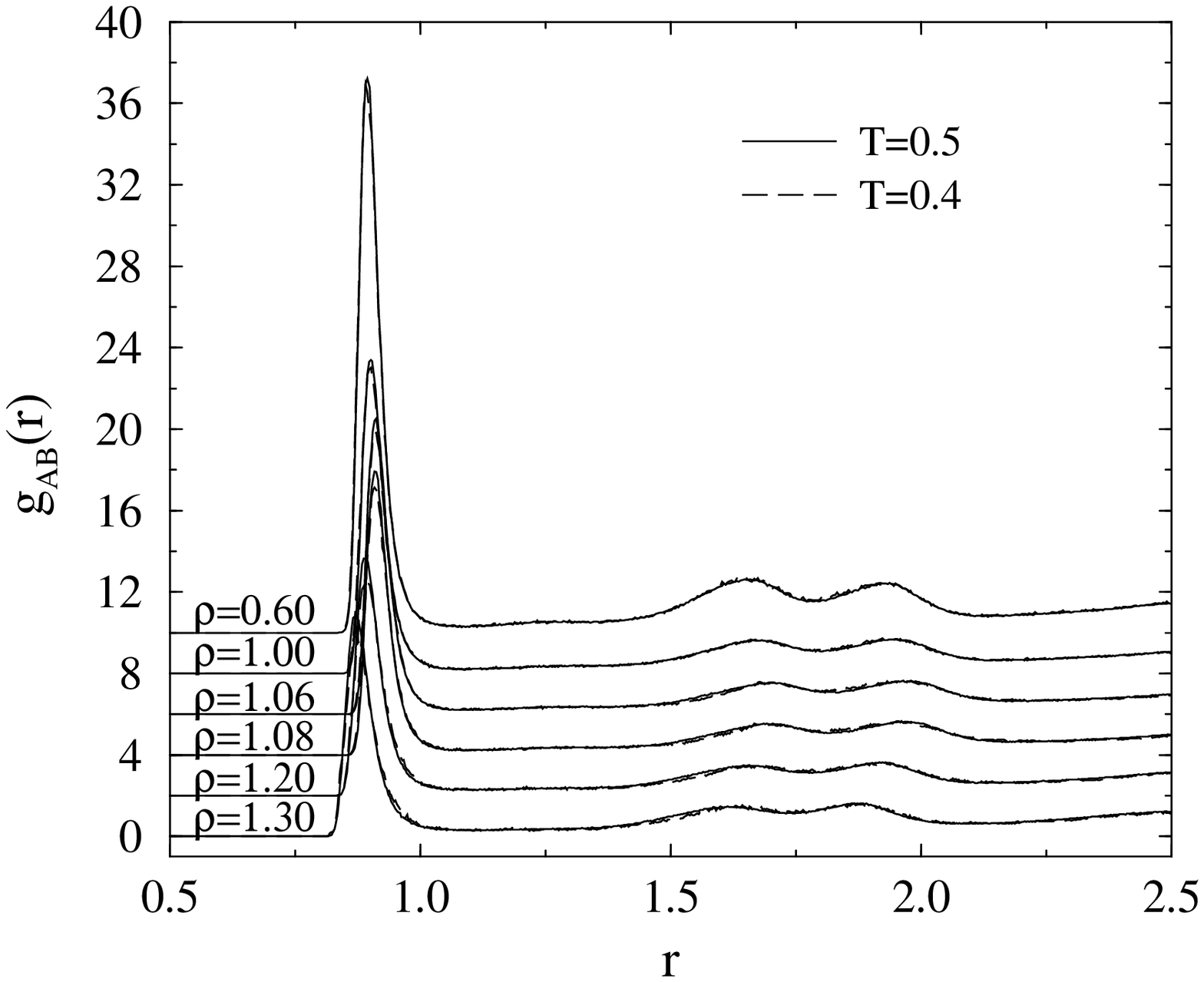}}
\caption{Partial radial distribution function for A particles (top)
and A and B particles (bottom)
averaged over reoptimised local minima from the samples
generated at $T=0.5$ and $T=0.4$. 
For clarity, the curves have been displaced vertically.}
\label{fig:rdfAAc}
\end{figure}

\begin{figure}
\begin{psfrags}
\psfrag{15}[cr][cr]{\Large 15}
\psfrag{12}[cr][cr]{\Large 12}
\psfrag{9}[cr][cr]{\Large 9}
\psfrag{6}[cr][cr]{\Large 6}
\psfrag{3}[cr][cr]{\Large 3}
\psfrag{0}[cr][cr]{\Large 0}
\psfrag{0.5}[tc][tc]{\Large 0.5}
\psfrag{1.0}[tc][tc]{\Large 1.0}
\psfrag{1.5}[tc][tc]{\Large 1.5}
\psfrag{2.0}[tc][tc]{\Large 2.0}
\psfrag{2.5}[tc][tc]{\Large 2.5}
\psfrag{T=0.5}[Bl][Bl]{$T=0.5$}
\psfrag{T=0.4}[Bl][Bl]{$T=0.4$}
\psfrag{r=0.60}[Bl][Bl]{$\rho=0.60$}
\psfrag{r=1.00}[Bl][Bl]{$\rho=1.00$}
\psfrag{r=1.06}[Bl][Bl]{$\rho=1.06$}
\psfrag{r=1.08}[Bl][Bl]{$\rho=1.08$}
\psfrag{r=1.20}[Bl][Bl]{$\rho=1.20$}
\psfrag{r=1.30}[Bl][Bl]{$\rho=1.30$}
\psfrag{gBB}[Bc][tc]{\Large $g_{\rm BB}(r)$}
\psfrag{r}[tc][tc]{\Large $r$}
\centerline{
\includegraphics[width=14cm]{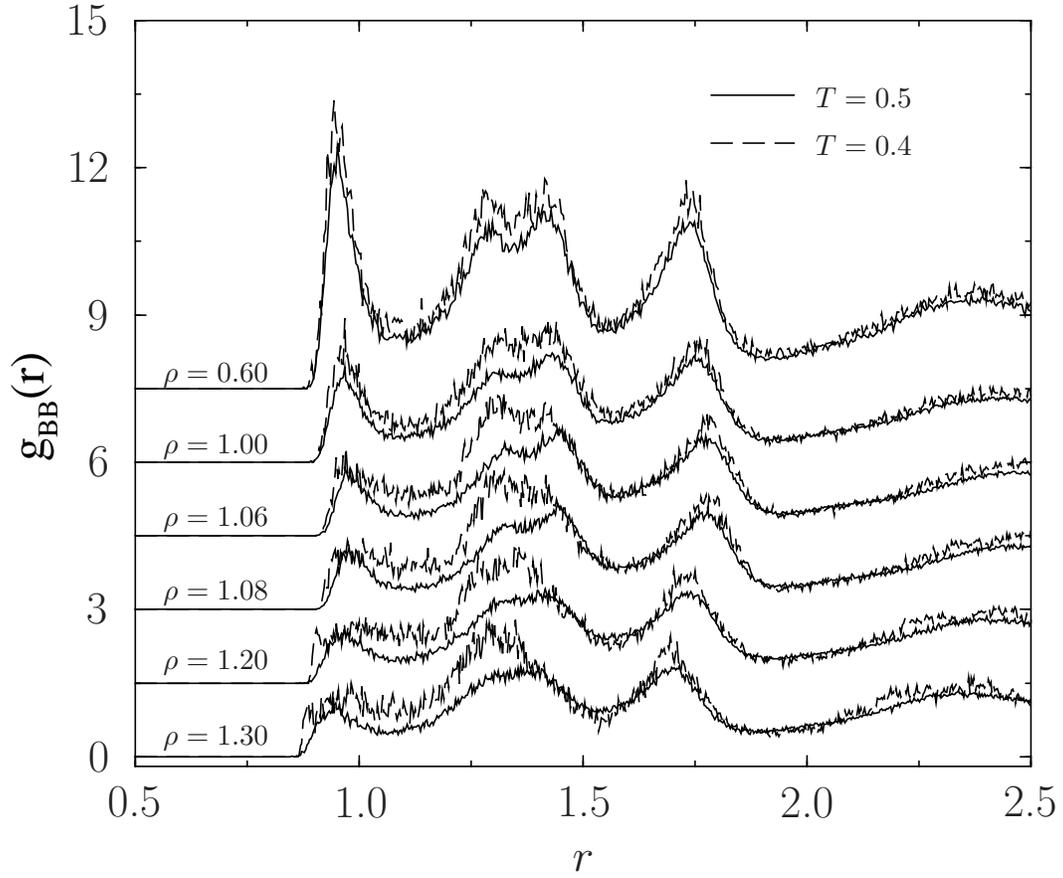}}
\end{psfrags}
\caption{Partial radial distribution function for B particles 
averaged over reoptimised local minima from the samples generated at
$T=0.5$ and $T=0.4$.
For clarity, the curves have been displaced vertically.}
\label{fig:rdfBBc}
\end{figure}

Sastry et al.\cite{Sastry97} suggested that the limiting
density, in our case $\rho_l=1.06$, 
is a lower limit to glass formation.
In order to elucidate this signature we compare the PRDF 
averaged over reoptimised local minima initially sampled at
$T=0.5$ and $T=0.4$.   
The results in Fig.~\ref{fig:rdfAAc} reveal a clear difference between 
the two samples for densities above $\rho_l=1.06$. 
For the lower temperature, corresponding to the glassy phase,
the structure is more ordered. However, below $\rho_l=1.06$ the PRDF 
are nearly the
same, so it is not possible to 
differentiate between the glass and supercooled liquid structures 
in this way.
Similar behaviour is exhibited by $g_{BB}$ in Fig.~\ref{fig:rdfBBc}. 
However, $g_{AB}$, illustrated in Fig.~\ref{fig:rdfAAc},
does not separate the two phases, indicating that the A-B 
disorder is similar in the glass and the supercooled liquid.   

\begin{figure}
\psfrag{r=1.20}[tc][tc]{\Large $\rho=1.20$}
\psfrag{r=1.08}[tc][tc]{\Large $\rho=1.08$}
\psfrag{r=1.06}[tc][tc]{\Large $\rho=1.06$}
\psfrag{r=1.04}[tc][tc]{\Large $\rho=1.04$}
\psfrag{r=1.02}[tc][tc]{\Large $\rho=1.02$}
\psfrag{r=1.00}[tc][tc]{\Large $\rho=1.00$}
\psfrag{r=0.98}[tc][tc]{\Large $\rho=0.98$}
\psfrag{r=0.96}[tc][tc]{\Large $\rho=0.96$}
\psfrag{r=0.60}[tc][tc]{\Large $\rho=0.60$}
\centerline{
\includegraphics[width=14cm]{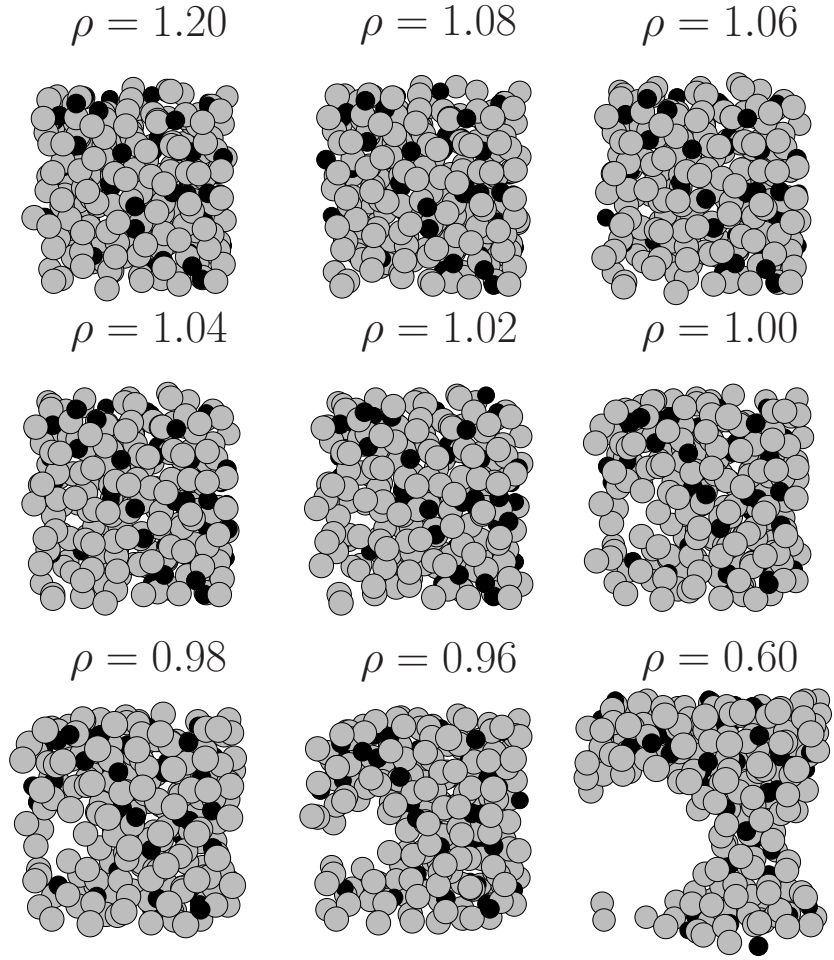}}
\caption{Perspective view of the A (gray) and B (black) atoms 
for one particular local minimum reoptimised at different 
densities and initially located in an MD run with
$T=1$ as the kinetic equipartition temperature.}
\label{fig:config_void}
\end{figure}

The formation of a void in one particular local
minimum sampled at $T=1$ is shown in Figure \ref{fig:config_void}.
The void starts to form at the limiting density 
$\rho_l=1.06$ and then increases when the density is reduced further. 



\section{Conclusions}
In this paper we have considered the effect of the density on local minima 
of a binary Lennard-Jones 
system sampled at kinetic equipartition 
temperatures $T=1.0$, $0.5$ and $0.4$.
A limiting density is found around $\rho_l=1.06$,  
below which fractures and voids 
begin to form. The latter features cause discontinuities in various 
properties, in agreement with 
Sastry's results.\cite{Sastry00} 

In the regime $\rho_l=1.06\leq \rho \leq 1.3$  
the main peaks in 
the PRDF for AA, BB, and AB 
particles are slightly displaced to longer distance relative to
$\rho=1.3$. Below $\rho_l$, fractures and voids 
begin to form and  the main peaks then move in the opposite
direction. In the PRDF for BB particles at low density, 
a clear splitting of the second peak appears as a consequence of a 
more ordered local structure. Furthermore, below $\rho_l$, the PRDF for
AA, BB and AB particles 
for the samples generated at $T=0.5$ and
$T=0.4$ are indistinguishable, so it is not possible to differentiate
between the glass and supercooled liquid structures at low density 
in this way.   


The lowest value of the mean pressure also
occurs around $\rho_l$ for each temperature.
Below this density, the mean potential energy 
changes almost linearly with $\rho$ due
to a transition from homogeneous local 
minima to inhomogeneous structures with voids.\cite{Sastry97} 
The minimum of the mean energy potential 
as a function of the density corresponds to a pressure close to zero.

On the other hand, at positive
pressure (compressed region), we found that the mean 
normal mode frequencies increase with energy but,
at negative pressure (stretched region), the opposite behaviour appears.
At lower density, 
the mean frequency is practically the same for 
samples obtained at each temperature,
and reaches a limiting value as the density decreases.

The initial MD trajectories sample fewer local minima at low temperature,
as expected.      

\section{Acknowledgements}
J.~H.-R.~ thanks Dr.~J.~P.~K.~Doye for numerous helpful discussions.
Dr.~Matt Hodges' program 
Xmakemol was used for visualising the positions of
the atoms in the system.
J.H.-R. also gratefully acknowledges the support of the 
Ministerio de Ciencia y Tecnolog\'{\i}a and FEDER under 
Grant No. BFM2001-3343.


\end{document}